\documentclass[conference]{IEEEtran}
\IEEEoverridecommandlockouts
\usepackage{cite}
\usepackage{amsmath,amssymb,amsfonts}
\usepackage{algorithmic}
\usepackage{graphicx}
\usepackage{textcomp}
\usepackage{xcolor}
\usepackage{url}
\usepackage{algorithm}
\usepackage{algorithmic}

\def\BibTeX{{\rm B\kern-.05em{\sc i\kern-.025em b}\kern-.08em
    T\kern-.1667em\lower.7ex\hbox{E}\kern-.125emX}}
\begin{document}

\title{Meta-Reinforcement Learning via Buffering Graph Signatures for Live Video Streaming Events}

\author{\IEEEauthorblockN{Stefanos Antaris}
\IEEEauthorblockA{\textit{KTH Royal Institute of Technology} \\
\textit{Hive Streaming AB}\\
Sweden \\
antaris@kth.se}
\and
\IEEEauthorblockN{Dimitrios Rafailidis}
\IEEEauthorblockA{\textit{University of Thessaly} \\
Greece \\
draf@uth.gr}
\and
\IEEEauthorblockN{Sarunas Girdzijauskas}
\IEEEauthorblockA{\textit{KTH Royal Institute of Technology} \\
Sweden \\
sarunasg@kth.se}
}

\maketitle

\begin{abstract}
In this study, we present a meta-learning model to adapt the predictions of the network's capacity between viewers who participate in a live video streaming event.  We propose the MELANIE model, where an event is formulated as a Markov Decision Process, performing meta-learning on reinforcement learning tasks. By considering a new event as a task, we design an actor-critic learning scheme to compute the optimal policy on estimating the viewers' high-bandwidth connections. To ensure fast adaptation to new connections or changes among viewers during an event, we implement a prioritized replay memory buffer based on the Kullback-Leibler divergence of the reward/throughput of the viewers' connections. Moreover, we adopt a model-agnostic meta-learning framework to generate a global model from past  events. As viewers scarcely participate in several events, the  challenge resides on how to account for the low structural similarity of different events. To combat this issue, we design a graph signature buffer to calculate the structural similarities of several streaming events and adjust the training of the global model accordingly. We evaluate the proposed model on the link weight prediction task on three real-world datasets of live video streaming events.  Our experiments demonstrate the effectiveness of our proposed model, with an average relative gain of $25\%$ against state-of-the-art strategies.  For reproduction purposes, our evaluation datasets and implementation are publicly available at \url{https://github.com/stefanosantaris/melanie}
\end{abstract}

\begin{IEEEkeywords}
Meta-reinforcement learning, graph signatures, video streaming
\end{IEEEkeywords}

\section{Introduction} \label{sec:introduction}

\begin{figure}[t!]

    \centering
    \includegraphics[scale=0.2]{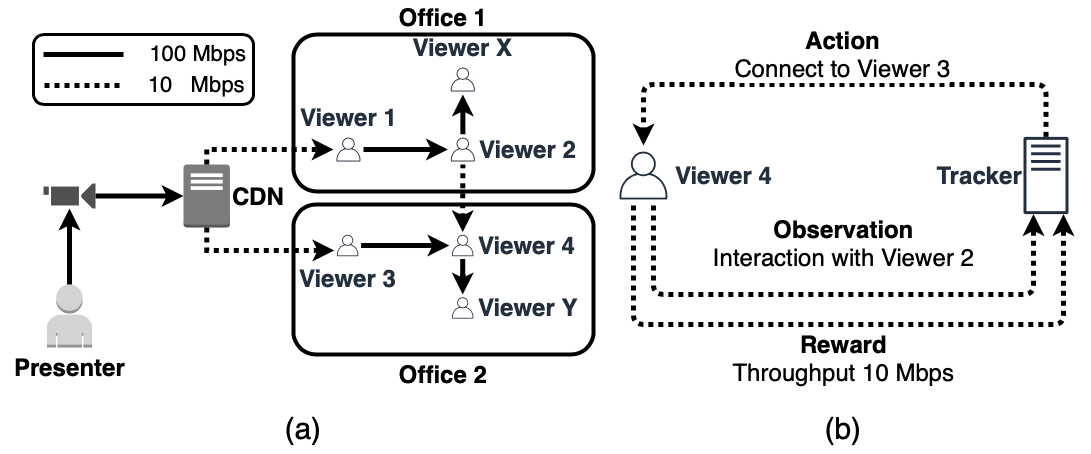}
    \caption{(a) A distributed live video streaming process in enterprise offices with limited network capacity. (b) Selection process of each viewer's connections via the tracker.}
    \label{fig:intro}
\end{figure}

Live video streaming services have been widely adopted by large enterprises to enable the communication among thousands of employees who are located at different offices across the world. Accounting for the high impact of video to the employee engagement and productivity, Fortune-500 companies converted more than $93\%$ of their physical to live video streaming events \cite{ibmadoptionreport}. However, network inefficiency and bandwidth congestion is experienced when several employees attend a high-quality live video streaming event at the same time. This negatively impacts the performance of the event, with more than $50\%$ of the viewers not attending the overall event~\cite{akamai}. To address this problem, large enterprises exploit distributed solutions provided by live video streaming companies e.g., Hive Streaming AB\footnote{\url{https://www.hivestreaming.com/}}. Such solutions leverage the office's internal high-bandwidth network to efficiently distribute the live video stream between viewers \cite{antaris2020vstreamdrls, Roverso2015}. For example, as shown in Figure \ref{fig:intro}(a), Viewers $1$ and $3$ directly connect to the Content Delivery Network (CDN) through a low-bandwidth connection to download the live video stream of the presenter. Viewers $1$ and $3$ exploit the internal high-bandwidth network to distribute the live video stream to Viewers $2$ and $4$, respectively. Then, Viewers 2 and 4 distribute the video stream to the remaining viewers at the same office, for example to Viewers $X$ and $Y$, accordingly.

To immediately select the internal high bandwidth connections, distributed video streaming solutions require prior knowledge of the company's network topology, for instance, Viewers $1$ and $2$ are in the same Office $1$. Such information is not always feasible to be acquired, since large enterprises regularly adapt their network \cite{antaris2020vstreamdrls}. Therefore, it is essential to predict the network capacity of each connection in real-time during the live video streaming event based on the observed viewers' interactions. To infer the high-bandwidth connections, each viewer periodically reports various data, such as throughput, interactions, and so on, to a centralized server, namely tracker. The tracker exploits the reported data to predict the network capacity among any two viewers of the event. Then, the tracker selects the connections with the highest predicted throughput of the viewers and adapts their connections. For example, as shown in Figure \ref{fig:intro}(b), the tracker probes Viewer $4$ to adjust the low-bandwidth connection ($10$ Mbps) with Viewer $2$ and connect to Viewer $3$ with a high-bandwidth connection ($100$ Mbps). The problem of selecting the proper connections among viewers becomes even more challenging at the beginning of the event, only after ``exploring'' a limited number of connections. To avoid network congestion, it is necessary to predict the best possible connection almost immediately. Nonetheless, this might only be possible by relying on the knowledge from the similar events that happened in the past on the same or similar networks.

The viewers' interactions during a video streaming event can be modeled as a temporal interaction network. The edge/ interaction weight of the network corresponds to the measured throughput of the connection among two nodes/viewers. Graph representation learning has been proven to be an efficient strategy to learn low-dimensional node embeddings that capture the evolution of the network~\cite{dygnn2020, jodie2019, ctdne2018, pareja2020evolvegcn, sankar2020dysat}. Baseline approaches exploit Recurrent Neural Networks (RNNs) \cite{jodie2019, pareja2020evolvegcn}, self-attention mechanisms \cite{sankar2020dysat} and Long-Short Term Memory (LSTMs)  units \cite{dygnn2020} to capture the network's evolution. Although evolving graph representation learning strategies can capture the temporal evolution of the graph, such strategies do not necessarily work on streaming events. At the beginning of a live video streaming event a viewer may have more low-bandwidth interactions than high-bandwidth connections~\cite{antaris2021gels}. Existing strategies learn to accurately predict the low weight edges/interactions and require a large number of interactions to estimate the high weights of the connections among viewers~\cite{jodie2019,antaris2020vstreamdrls}. The challenge here is how to accurately predict the high weights of the edges/interactions, given only a few interactions per viewer. Recently, graph representation learning approaches have been proposed to predict the high weights of edges/interactions in a live video streaming event \cite{antaris2020vstreamdrls, antaris2020egad}. However, these approaches exploit the viewers' interactions in a single event, ignoring the knowledge of past live video streaming events.

Extracting knowledge from past streaming events is a challenging task, as a significantly low number of viewers attends multiple live video streaming events, as it will be demonstrated in Section I of our supplementary~\cite{hyper_sup}. In addition, each event has different network characteristics such as edges/interactions weight distribution, viewers' emerging patterns, and so on. Therefore, each live video streaming event is considered as a new task, which the prediction model needs to adapt. \textit{Meta-Learning} has achieved remarkable performance on computing generic models, able to adjust to new tasks when only a limited amount of information is available \cite{bose2020metagraph, finn2017maml, Lu2020metahin}. The goal of meta-learning is to derive a global knowledge from previous experiences, so as to rapidly adapt to a new task with only a small amount of training data. Meta-learning approaches have successfully been employed on various machine learning tasks, such as image classification \cite{siavash2019cactu, Zhu2020, Wei2019} and recommender systems \cite{Lu2020metahin, pan2019warm, NIPS2017_51e6d6e6}. However, meta-learning on graphs, where each task corresponds to a new graph, has received little attention. Meta-Graph adopts the Model-Agnostic Meta-Learning framework \cite{finn2017maml} and graph signature functions \cite{brockschmidt2020gnn} to perform link prediction on new graphs with limited edges \cite{bose2020metagraph}. Nonetheless Meta-Graph works on static graphs, and does not reflect on the dynamic case of live video streaming events. Recently, GELS employed gradient boosting to extract information from past streaming events for improving user experience~\cite{antaris2021gels}. However, GELS considers equal contribution of each event when learning the global model. Provided that each event exhibits different characteristics, it is essential to compute the similarity of different events and generate an accurate global model.

In this paper, we propose a {M}eta-r\textbf{E}inforcement \textbf{L}e\textbf{A}r\textbf{N}ing model via buffering graph s\textbf{I}gnatur\textbf{E}s, namely MELANIE, making the following contributions:
% . We formulate a live video streaming event as a Markov Decision Process (MDP), and employ deep Reinforcement Learning (RL) to estimate the high-bandwidth connections. We perform meta-learning to past events and generate a global model that exhibits fast adaptation to a new event within a limited number of viewers' interactions. Our main contributions are summarized as follows: 
i) \textit{ We implement a task adaption component to update the global MELANIE model for a new streaming event. We adopt the Actor-Critic reinforcement learning scheme, where the agent/tracker learns an optimal policy by incorporating the viewers' interactions. To quickly adapt to a new event, we design a replay memory buffer and prioritize the stored interactions based on the viewers' throughput divergence. Hence, the agent/tracker is trained on diverse experiences and converges to the optimal policy after a few interactions.; ii) We propose a meta-learning component to generate the global MELANIE model by combining the learned policy by the task adaptation component of the current live video streaming event and the previously trained events. We formulate two objective functions to update the actor and critic parameters, respectively, based on the structural similarities of different streaming events; iii) We design a graph signature memory buffer in the meta-learning component to compute the network's structure similarity between different events. In doing so, we enforce to produce similar representations of viewers with high structural similarities in different events. The graph signature memory buffer allows the meta-learning component to generate the global MELANIE model by generalizing over several events with a few common viewers.}

Our experiments on three real-world datasets, generated by live video streaming events, demonstrate that superiority of MELANIE over several baseline approaches.
The remainder of the paper is organized as follows, in Section \ref{sec:preliminaries} we formally define the problem of meta-reinforcement learning in live video streaming events and detail the proposed MELANIE model in Section \ref{sec:proposed_model}. Our experimental evaluation is presented in Section \ref{sec:experimental}, and we conclude the study in Section \ref{sec:conclusions}.

\begin{figure*}[t!]

    \centering
    \includegraphics[scale=0.18]{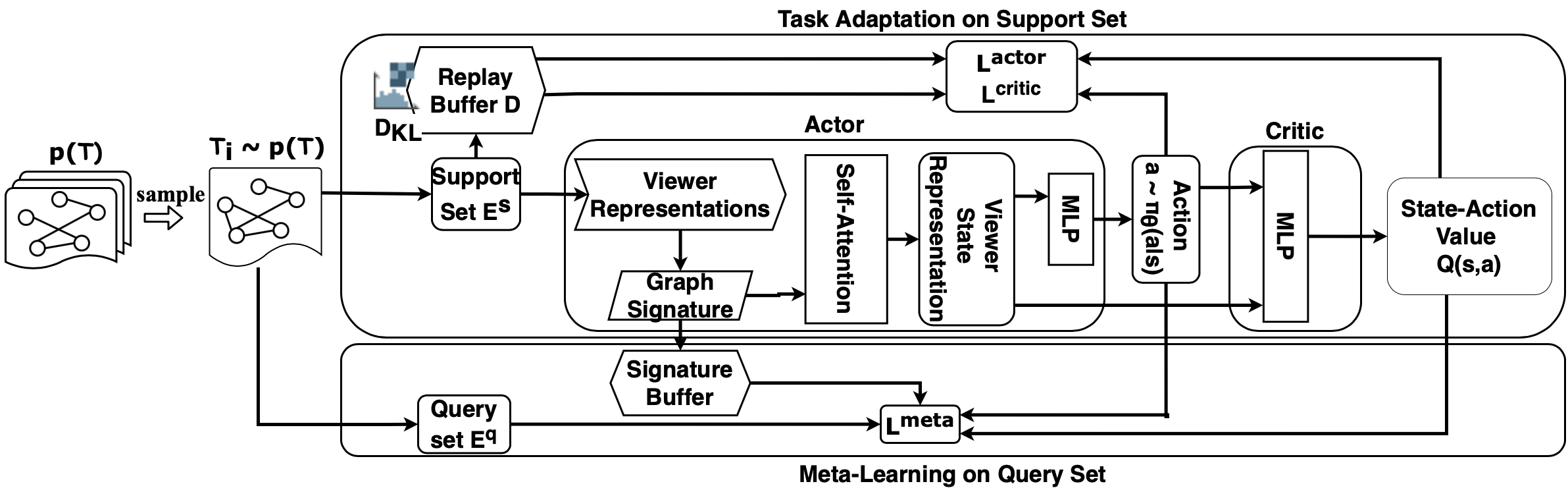}
    \caption{Overview of the MELANIE model, given a new task, that is a new temporal interaction network $p(\mathcal{T})$. MELANIE consists of (i) the task-adaptation component on the support set $\mathcal{E}^s$ and (ii) the meta-learning component on the query set $\mathcal{E}^q$.}
    \vspace{-0.2cm}
    \label{fig:local_framework}
\end{figure*}

\section{Preliminaries} \label{sec:preliminaries}

A distributed live video streaming event is represented as a temporal interaction network, which is defined as follows \cite{jodie2019, dygnn2020, ctdne2018}.

\textbf{\textit{Temporal Interaction Network}}. A temporal interaction network is defined as a graph $\mathcal{G} = (\mathcal{V}, \mathcal{E})$, with $N$ nodes/viewers $\mathcal{V} = \{ v_1, v_2,\ldots,v_N \}$ and an ordered sequence of $T$ node/viewer interactions/connections $\mathcal{E} = \{e^1, e^2, \ldots, e^T\}$. An interaction/connection $e^t_{u,v} = (u, v, t, b^t_{u,v})$ occurs between two nodes/ viewers $u\in\mathcal{V}$ and $v\in\mathcal{V}$ at time $t \in \mathbb{R^+}$, $0 \leq t \leq T$. Each interaction/connection $e^t_{u,v}$ has a weight $b^t_{u,v} \in \mathcal{B}^t$, which corresponds to the throughput measurement of node/viewer $u$, reported in the tracker. The set $\mathcal{B}^t$ contains the network's weights at time $t$.

The nodes/viewers periodically communicate with the tracker to retrieve the interactions/connections, established during a live video streaming event (Section \ref{sec:introduction}). Therefore, the tracker is responsible to determine the viewers' edges/connections. We model the selection of the node/viewer connections/interactions during a live video streaming event as a Markov Decision Process (MDP). 

\textbf{\textit{Live Video Streaming MDP}}
A MDP is defined as $\mathcal{M} = (\mathcal{S}, \mathcal{A}, \mathcal{P}, \mathcal{R}, \gamma)$, with $\mathcal{S}$, $\mathcal{A}$, $\mathcal{P}$,  $\mathcal{R}$ and $\gamma$ being the state, action, transition probability and reward sets, and the discount factor, respectively. At each time step $t \in \mathbb{R}^+$, the tracker/agent takes an action $a^t \in \mathcal{A}$ to connect/interact node/viewer $u\in\mathcal{V}$ with node $v \in \mathcal{V}$, based on state $s^t \in \mathcal{S}$ of the node/viewer $u \in\mathcal{V}$. The tracker receives the measured throughput $b^t_{u,v}$ of the interaction $e^t_{u,v}$ as a reward $R(s^t,a^t) \in \mathcal{R}$, and the node/viewer $u$'s state is updated to $s^{t+1}$ with transition probability $p(s^{t+1}| s^t,a^t) \in \mathcal{P}$. The objective of the tracker/agent is to find an optimal policy $\pi_{\theta}: \mathcal{S}\times\mathcal{A} \rightarrow [0,1]$, where $\theta$ are the parameters of the policy $\pi$, that maximizes the expected cumulative rewards from any state-action pairs as follows:
$$\mathcal{Q}^{*}(s,a) = \max_{\pi_{\theta}} \mathbb{E}_{\pi_{\theta}}\{\sum_{k=0}^{\infty}\gamma^{k}R(s,a) | s= s^t, a=a^t\}$$ where $\mathbb{E}_{\pi_{\theta}}$ is the expectation based on the policy $\pi_{\theta}$, $\gamma$ is the discount factor, and $R(s,a) \in \mathcal{R}$ is the reward received at the time step $t$, given the action $a^t$ taken at the state $s^t$.

% Add definition
Finding the optimal policy $\pi_{\theta}$ of the agent/tracker in a live video streaming event typically requires a large number of node/viewer interactions to achieve efficient performance. Our study is inspired by the gradient based meta-learning framework in reinforcement learning (meta-RL), which optimizes the global policy $\pi_{\theta}$ of the agent over several tasks, so as to rapidly adapt to a new task \cite{hochreiter2001learning, finn2017maml}. Extracting knowledge from live video streaming events is defined as the following meta-RL process:

\textbf{\textit{Meta-RL in Live Video Streaming Events}}
In meta-RL for live video streaming events, we consider a distribution of tasks $p(\mathcal{T})$, where each task corresponds to a live video streaming event. We denote each task by $\mathcal{T}_i = (\mathcal{M}_i, \mathcal{G}_i)$, where $\mathcal{M}_i = (\mathcal{S}, \mathcal{A}, \mathcal{P}, \mathcal{R}, \gamma)$ and $\mathcal{G}_i = (\mathcal{V}_i, \mathcal{E}_i)$ is the MDP and the temporal interaction network of the live video streaming event, respectively. During meta-training, for each task $\mathcal{T}_i \sim p(\mathcal{T})$ we divide the node/viewer interactions $\mathcal{E}_i$ in a support set $\mathcal{E}^s_i \subseteq \mathcal{E}_i$ and a query set $\mathcal{E}^q_i \subseteq \mathcal{E}_i$, with $\mathcal{E}^s_i \cap \mathcal{E}^q_i = \emptyset$. The agent/tracker adapts the policy $\pi_{\theta}$ to learn the task $\mathcal{T}_i$ based on the loss on the support set $\mathcal{E}^s_i$. Then, the meta-learner exploits the query set $\mathcal{E}^q_i$ to optimize the global policy $\pi_{\theta}$ across all tasks $p(\mathcal{T})$. The goal of meta-RL is to learn a global policy $\pi_{\theta}$ to maximize the following expected cumulative reward $$\displaystyle\max_{\theta}\displaystyle\sum_{\mathcal{T}_i  \in p(
\mathcal{T})}\mathbb{E}_{\pi_{\theta}}\{\sum_{k=0}^{\infty}\gamma^{k}R(s,a) | s= s^t, a=a^t\}$$

\section{Proposed Model} \label{sec:proposed_model}

\subsection{Overview of MELANIE}

As illustrated in Figure \ref{fig:local_framework}, the proposed MELANIE model consists of two main components: the task adaptation and the meta-learning component. The goal of the proposed MELANIE model is to learn a generic model that achieves fast adaptation to new MDPs, that is new live video streaming events.

\textbf{Task Adaptation} The role of this component is to adapt the global policy $\pi_{\theta}$ of the tracker to a new task $\mathcal{T}_i \sim p(\mathcal{T})$. The input of the task adaptation component is the support set $\mathcal{E}^s_i$ of the sampled task $\mathcal{T}_i$. We adopt an Actor-Critic reinforcement learning scheme to model the interactions between viewers and the tracker. The policy $\pi_{\theta}$ of the tracker is trained based on the state-action transitions stored in a replay memory buffer. Following \cite{schaul2015prioritized}, we prioritize the state-action transitions in the replay memory buffer based on the KL-divergence of the viewers' reward distribution between consecutive time steps. Provided that most viewer interactions may have low reward at the beginning of the live video streaming event, we train the policy $\pi_{\theta}$ on dissimilar experiences.

\textbf{Meta-learning} The role of the meta-learning component is to evaluate the policy $\pi_{\theta}$, learned by the task adaptation component, to the query set $\mathcal{E}^q_i$ of the sampled task $\mathcal{T}_i$. To measure the similarity of the sampled task $\mathcal{T}_i$ against the previously trained tasks, we introduce a signature buffer during the meta-learning process. The signature buffer contains a signature of each task $\mathcal{T}_i \sim p(\mathcal{T})$, which represents the structural information of the task's network. According to the gradient-based Model Agnostic meta-learning approach~\cite{finn2017maml}, we update the global policy $\pi_{\theta}$ to generate the global MELANIE model over all the computed signatures of the tasks.

\subsection{Task Adaptation on the Support Set} \label{sec:local_model}

The input of the task adaptation component is the support set $\mathcal{E}^s_i$ of the sampled task $\mathcal{T}_i \sim p(\mathcal{T})$. Given the viewers' interactions of the support set $\mathcal{E}^s_{i}$, we aim to learn the optimal policy $\pi_{\theta}$ of the specific task $\mathcal{T}_i$, by adopting a deep RL framework based on the Actor-Critic learning scheme. Therefore, the task adaptation component consists of three main parts: i) the Actor network, ii) the Critic network, and iii) the replay memory buffer.

\textbf{Actor Network} At each time step $t$, the actor network takes as input an interaction $e^t_{u,v} \in \mathcal{E}^s_i$ and then generates an action $a^t$ for the node/viewer $u \in \mathcal{V}_i$ of the task $\mathcal{T}_i \sim p(\mathcal{T})$. We represent each node/viewer $u\in \mathcal{V}_i$ with a $d$-dimensional node/viewer embedding $\mathbf{X}_u \in \mathbb{R}^d$ based on the node/viewer features~\cite{Lee2019}. If the nodes/viewers do not have any node features, as in the case of live video streaming events, we calculate an embedding look up. Given an interaction $e^t_{u,v} \in \mathcal{E}^s_i$, we compute a $d_s$-dimensional state representation $s^t_u \in \mathbb{R}^{d_s}$ of the viewer $u \in \mathcal{V}_i$, as follows \cite{gat2018, brockschmidt2020gnn}:

\begin{equation}
\mathbf{s}^t_u = g_{\phi}(u, \mathcal{N}^t_u) = \sigma\bigg( \displaystyle \sum_{v \in \mathcal{N}^t_u}  c_{u,v} \mathbf{W}^S \mathbf{X}_u\bigg)
\end{equation}
where $g_{\phi}(u, \mathcal{N}^t_u)$ is the attention function of the node/viewer $u \in \mathcal{V}_i$ with its neighborhood $\mathcal{N}^t_u = \{v: e^t_{u,v} \in \mathcal{E}^t_{u}\}$, parameterized by $\phi = \{\mathbf{W}^S\}$. The parameter weight matrix $\mathbf{W}^S \in \mathbb{R}^{d_s \times d}$ is the shared weight transformation of the node/viewer embedding $\mathbf{X}_u$ to a $d_s$-dimensional vector and $\sigma(\cdot)$ is the exponential linear unit (ELU) activation function. The attention coefficient $c_{u,v}$ measures the similarity of the node/viewer $v$ to the node/viewer $u$ as follows:

\begin{equation}
    \begin{array}{c}
         c_{u,v} = \frac{exp(o_{u,v})}{\displaystyle \sum_{w \in \mathcal{N}^t_u} exp(o_{u,w})} \\
         o_{u,v} = \delta\bigg( b^t_{u,v} \cdot \mathbf{H}^{T}[\mathbf{W}^{S} \mathbf{X}_u || \mathbf{W}^{S} \mathbf{X}_v] \bigg)
    \end{array}
\end{equation}
where $b^t_{u,v}$ is the connection weight/throughput between nodes/ viewers $u$ and $v$ at the time step $t$, $||$ is the concatenation of the node/viewer representations $\mathbf{X}_u$ and $\mathbf{X}_v$, and $\delta(\cdot)$ is the LeakyRELU non-linearity activation function. Provided that the temporal interaction network $\mathcal{G}_i$ of the sampled task $\mathcal{T}_i$ evolves over time, we capture the evolution in the state representation $s^t_u$, by parameterizing the attention function with a $2d_s$-dimensional graph signature $\mathbf{H} \in \mathbb{R}^{2d_s}$, which is defined as follows:

% We parameterize the attention function with a graph signature $\mathbf{H} \in \mathbb{R}^{2d_s}$, defined as follows:

\begin{equation}
    \mathbf{H} = \psi_{\beta}(\mathbf{X}) = \sigma(\mathbf{W^H}\mathbf{X} + \mathbf{b^H})
\label{eq:signature}
\end{equation}
where $\psi_{\beta}$ is the aggregation of the node representations $\mathbf{X}$ to a $N$-dimensional signature vector $\mathbf{H}$, parameterized by $\beta = \{\mathbf{W^H}, \mathbf{b^H}\}$. Intuitively, the role of the graph signature $\mathbf{H}$ is to capture the structural and node/viewer features similarities of the temporal interaction network over time, and bias the node/viewer feature aggregation towards nodes/viewers with similar interactions. By computing the graph signature $\mathbf{H}$, we determine the structural similarity of viewers, and promote similar viewers in the attention process \cite{brockschmidt2020gnn}.

The state $\mathbf{s}^t_u$ captures the preferences of the viewer $u$ at the $t$-th time step. An interaction $e_{u,v}$ with high connection throughput corresponds to a high attention coefficient $c_{u,v}$, which reflects on the preferences of the viewer $u$ in the state representation $\mathbf{s}^t_u$. The actor network transforms the state representation  $s^t_u$ of the viewer $u \in \mathcal{V}_i$ to an $N$-dimensional action $\mathbf{a}^t$ vector. In our implementation, we employ a two-layer perceptron (MLP) on the state representation $s^t_u$ as follows:

\begin{equation}
    \mathbf{a}^t = \pi_{\theta}(s^t) = MLP(s^t)
\label{eq:action}
\end{equation}
We normalize the action vector $\mathbf{a}^t$ based on the softmax function and select the node/viewer with the highest value. 
%Note that the agent's policy $\pi_{\theta} = \{g_{\phi}, \psi_{\beta}\}$ consists of the attention function $g_{\phi}$, the linear modulation function $\psi_{\beta}$ and the action representation function, parameterized by the weight parameters $\theta = \{\phi, \beta, \mu\}$.
We employ the $\epsilon$-greedy exploration technique to learn an accurate policy $\pi_{\theta}$ \cite{sutton2018}. 

\textbf{Critic Network} The input of the Critic network is the node/ viewer state $s^t_u$ and the action $a^t$ generated by the policy $\pi_{\theta}$. The critic network outputs a scalar, which is an approximation of the true state-action value function $Q_{w}(s^t,a^t)$, that is the following Q-value function:

\begin{equation}
    Q_w(s^t_u,a^t) = f_{w}(s^t_u,a^t) = MLP(s^t_u \oplus a^t) 
\label{eq:q_value}
\end{equation}
where $\oplus$ denotes the concatenation of the state representation $s^t_u$ and the action representation $a^t$. Here $f_{w}$ is the Q-value approximation function paremeterized by $w$, which consists of the weights and biases in the MLP. The Q-value function represents the benefit of the action $a^t$ generated by the Actor network, given the node/viewer state $s^t_u$ and the policy $\pi_{\theta}$. Following the deterministic policy gradient theorem \cite{silver2014}, we update the parameters $\theta$ of the Actor based on the sampled policy gradient as \cite{sutton2018}:

$$
     \theta \leftarrow \theta - \eta \nabla_{\theta} \mathcal{L}^{actor}(\pi_{\theta}; \mathcal{E}^s_i) 
$$
\begin{equation}     
     \begin{array}{l}
      \mathcal{L}^{actor}(\pi_{\theta}; \mathcal{E}^s_i) =\\  -\frac{1}{K} \sum_{t=0}^{K} log \pi_{\theta} (a| s)[R(s, a) - Q_w(s,a)]|_{s=s^t, a=\pi_{\theta}(s^t)}
\end{array}
\label{eq:loss_actor}
\end{equation}

\vspace{0.1cm}

\noindent where $R(s, a) - Q_w(s,a)$ indicates the merit of the action $a$, given the state $s$, compared with the random action, and $\eta$ is defined as the step size. The term $K$ is the number of interactions exploited to train the actor network. The weight parameters $w$ of the Critic network are updated accordingly via the temporal-difference learning approach, that is minimizing the following mean squared error:

\begin{equation}
\begin{array}{c}
     w \leftarrow w - \eta \nabla_{w} \mathcal{L}^{critic}(f_w; \mathcal{E}^s_i) \\ \\
     \mathcal{L}^{critic}(f_w; \mathcal{E}^s_i) = \frac{1}{K} \sum_{t=0}^{K}(R(s,a) - Q_{w}(s,a))^2 |_{s=s^t, a=\pi_{\theta}(s^t)}\\
\end{array}
\label{eq:loss_critic}
\end{equation}

\textbf{Replay Memory Buffer} To optimize the Actor and the Critic networks, we employ a replay memory buffer $\mathcal{D}$, that contains the $D=|\mathcal{D}|$ latest state-action transitions. The agent/tracker adapts the nodes/viewers connections based on their reward until the viewers weight/throughput distribution converges. Therefore, we store the $D$ latest state-action transitions of the nodes/viewers in the replay memory buffer $\mathcal{D}$. The replay memory buffer allows the agent/tracker to learn from earlier memories, and break undesirable temporal correlations, that is similar state-action transitions in a short period of time. Instead of exploiting all the state-action transitions to train the agent, we sample a subset of the stored transitions. The most popular sampling strategy is uniform sampling, where each transition in the replay memory buffer is selected with equal probability \cite{pmlr-v119-fedus20a, schaul2015prioritized}. However, provided that in our case the connections at the beginning of a live video streaming event may be among viewers with a low network's capacity, the majority of the stored state-action transitions in the replay memory buffer have low reward values. Hence, uniformly sampling the state-action transitions to train the Actor and Critic network prevents the agent/tracker to learn from new experiences. To overcome this problem, we prioritize the state-action transitions based on the Kullback-Leibler (KL) divergence\ values $D_{KL} (\mathcal{B}^t||\mathcal{B}^{t-1})$ of the weight/throughput distribution of each viewer between consecutive time steps. In particular, to measure the difference between the weight/throughput distributions $\mathcal{B}^t$ and $\mathcal{B}^{t-1}$ of two consecutive steps, the Kullback-Leibler (KL) divergence \cite{kullback1951information} is defined as follows:
$$D_{KL}(\mathcal{B}^t||\mathcal{B}^{t-1}) = \sum_{x \in \mathcal{X}} \mathcal{B}^t(x)log\bigg(\frac{\mathcal{B}^t(x)}{\mathcal{B}^{t-1}(x)} \bigg)$$
where $\mathcal{X}$ is the equally partitioned probability space of the interactions weight/throughput. A high KL-divergence value corresponds to significant changes on the weight/throughput distributions between consecutive live video streaming minutes, whereas a low value indicates that viewers converge to their optimal connections. We retrieve the top-$K \ll D$ state-action transitions from the buffer $\mathcal{D}$ and train the tracker/agent based on Equations \ref{eq:loss_actor} and \ref{eq:loss_critic}. 
%As we will demonstrate in Section \ref{sec:settings}, the sampling size $K$ of the replay memory buffer plays a crucial role in the performance of the task adaptive component.

\subsection{Meta-Learning on the Query Set} \label{sec:meta_model}

The goal of the meta-learning component is to learn the global weight parameters $\theta$ and $w$, so that the actor and critic networks  can quickly adjust to a novel task $\mathcal{T}$. The input of the meta-learning component is the query set $\mathcal{E}^q_i$ of the sampled task $\mathcal{T}_i \sim p(\mathcal{T})$ and the weight parameters $\theta$ and $w$ calculated by the task adaptation component. According to the gradient-based model of agnostic meta-learning~\cite{finn2017maml}, we update the weight parameters $\theta$ and $w$ on the query set $\mathcal{E}^q_i$ as follows:

\begin{equation}
    \begin{array}{c}
         \theta \leftarrow \theta - \eta \nabla_{\theta}\mathcal{L}^{meta}_{\theta}(\pi_{\theta}; \mathcal{E}^q_i)  \\ \\ 
         w \leftarrow w - \eta \nabla_{w}\mathcal{L}^{meta}_{w}(f_w; \mathcal{E}^q_i)
    \end{array}
\label{eq:meta_update}
\end{equation}
where $\eta$ is the learning rate. We formulate the loss functions $\mathcal{L}^{meta}_{\theta}$ and $\mathcal{L}^{meta}_{w}$ as follows:
\begin{equation}
\begin{array}{c}
    \mathcal{L}^{meta}_{\theta}(\pi_{\theta}; \mathcal{E}^q_i) = \mathcal{L}^{actor}(\pi_{\theta}; \mathcal{E}^q_i) - \frac{1}{|\mathcal{J}|}\sum_{j\in \mathcal{J}} KL(\mathbf{H}_i||\mathbf{H}_j) \\ \\

    \mathcal{L}^{meta}_{w}(w; \mathcal{E}^q_i) = \mathcal{L}^{critic}(f_{w}; \mathcal{E}^q_i) - \frac{1}{|\mathcal{J}|}\sum_{j\in \mathcal{J}} KL(\mathbf{H}_i||\mathbf{H}_j)
\end{array}
\label{eq:meta_loss}
\end{equation}
where the loss functions $\mathcal{L}^{actor}(\pi_{\theta}; \mathcal{E}^q_i)$ and $\mathcal{L}^{critic}(f_{w}; \mathcal{E}^q_i)$ measure the accuracy of the actor and the critic networks on the query set $\mathcal{E}^q_i$ based on  Equation \ref{eq:loss_actor} and \ref{eq:loss_critic}, respectively. The term $(\mathbf{H}_i||\mathbf{H}_j)$ is the KL divergence among different graph signatures. The intuition behind this divergence is to provide a model with parameters $\theta$ and $w$ and generalize over all the previously trained graph signatures $H_j \in \mathcal{J}$.

\textbf{Graph Signature Buffer} The graph signature buffer $\mathcal{J}$  contains the $C$ signatures $\mathbf{H}_i$ of each task $\mathcal{T}_i \sim p(\mathcal{T})$ that has been previously trained in the past events. Given the graph signature buffer, we compute the similarity among the newly sampled task $\mathcal{T}_i$ and the $C$ previous trained tasks. During the update of the parameters $\theta$ and $w$, we minimize the average divergence of the signature $\mathbf{H}_i$ of task $\mathcal{T}_i$ and all the stored graph signatures $\mathbf{H}_j \in \mathcal{J}$. Provided that different live video streaming events have a low number of common viewers, we capture the similarities between the events and promote similar viewers that attend to more than one event. In doing so, as we will show in the experimental evaluation, the meta-learning component can generalize in several divergent live video streaming events and quickly adjust to new events within a limited number of viewers' interactions. Further details about the proposed MELANIE algorithm can be found in \cite{hyper_sup}.

% In Algorithm \ref{algorithm:melanie}, we present the steps of the MELANIE model to learn a global policy $\pi_{\theta}$ of the agent/tracker. The input of MELANIE is a distribution of tasks $\mathcal{T}$, where each task corresponds to a streaming event, and the outputs are the global policy $\pi_{\theta}$ of the Actor network and the global parameters $f_w$ of the Critic network. We randomly initialize the parameters $\theta$ and $w$. In lines 3-10, we train our model to adapt to a new task $\mathcal{T}_i \sim \mathcal{T}$, given a set of support interactions $\mathcal{E}^s_i$. In line 4, we compute the action $a^t$ based on the policy $\pi_{\theta}$, followed by the Actor network and the state $s^t$ of the user $u$ in the interaction $e^t_{u,v}$. In line 5, the Critic network evaluates the selected action $a^t$ of the Actor network.  In Lines 6-9, the state-action transition is stored in the replay memory buffer $\mathcal{D}$ and  the $K$ most dissimilar state-actions are retrieved to update the parameters $\theta$ and $w$ based on  Equations \ref{eq:loss_actor} and \ref{eq:loss_critic}, respectively. At the end of the task adaptation component, in Lines 11-12 we compute the graph signature $\mathbf{H}_i$ of the task $\mathcal{T}_i$  and store it in the graph signature buffer $\mathcal{J}$. In lines 13-14, we update the global parameters $\theta$ and $w$ based on the query set $\mathcal{E}^q_i$ and Equation \ref{eq:meta_loss}. We repeat the process in lines 1-15 until we have experienced all the input tasks in $\mathcal{T}$.

\section{Experimental Evaluation} \label{sec:experimental}
\begin{table}[h]
	\caption{Summary statistics of  the three datasets.}
	\vspace{-0.3cm}
    \centering
    \resizebox{.45\textwidth}{!}{%
    \begin{tabular}{c|ccc}
        \hline
         \textbf{Datasets} & \textbf{LiveStream-1} & \textbf{LiveStream-2} & \textbf{LiveStream-3} \\ \hline
         \hline
         \#Events & 30 & 30 & 30 \\ \hline
         \#Offices & 62 & 12 & 56 \\ \hline
          \#Enterprises & 1 & 1 & 15 \\ \hline
         \#Viewers (K) & $24.752$ & $28.879$ & $148.972$ \\ \hline
         \#Connections (M) & $1.807$ & $0.590$ &  $4.139$ \\ \hline
         Avg. \#Events Per Viewer & $2.722 \pm 2.061$ & $1.349 \pm 0.883$ & $1.141 \pm 0.348$ \\ \hline
        %  AVG. Nodes  (K) & $2.243 \pm 1.633$ & $1.299 \pm 0.851$ & $6.297 \pm 5.386$ \\ \hline
        %  Average number of Interactions (K) & $0.599 \pm 0.700$ & $0.446 \pm 0.552$ & $ 0.389 \pm 0.582$ \\ \hline
        %  Average Viewer Participation & $2.722 \pm 2.061$ & $1.349 \pm 0.883$ & $1.141 \pm 0.348$ \\  \hline
         
    \end{tabular}
    }
    
    \label{tab:data_stats}
\end{table}

\subsection{Setup}

\textbf{Datasets} In our experiments, we evaluate our proposed MELANIE model on three real-world datasets~\cite{antaris2021gels}. In Table~\ref{tab:data_stats}, we summarize the statistics of each dataset and we perform an extensive analysis in Section I of our supplementary~\cite{hyper_sup}.

\textbf{Support/Train and Query/Test Sets} In our experiments, we evaluate our proposed MELANIE model in the link weight prediction task on the three examined datasets, with 30 temporal interaction networks (events). For each dataset, we consider $26$ live video streaming events as training set, $2$ for validation and $2$ for testing. Following \cite{Lu2020metahin}, we divide the interactions of each live video streaming event to support set $\mathcal{E}^s$ and query set $\mathcal{E}^q$ according to the interaction time, with a ratio of $8$:$2$. The task of link weight prediction is to predict the weight $b_{u,v}$ of the interaction $e_{u,v} \in \mathcal{E}^q$ that will occur in the query set $\mathcal{E}^q$ of the test set.

% we train our model o examine our meta-learning process In this paragraph, we will describe the evaluation protocol used to compare our proposed model against the baseline approaches. We will describe this as a link prediction task. First, we will remind (with reference to the Section \ref{sec:data_analysis}) that we have three datasets, with 30 graphs per dataset. We will mention that we use the first $26$ graphs for training, $2$ of the rest as validation and the final $2$ as testing. We divide the interactions of each graph with a ratio $80$:$20$ of support/query sets. We emphasize that during training, we keep the $80\%$ of the training interactions to adapt the model towards the specific task, and we generate the global model based on the rest $20\%$ of the interactions. We further mention that during testing, we similarly adapt the network towards the testing task with the $80\%$ of the interactions and we evaluate the performance of our model with the rest $20\%$ of the tasks.

\begin{table*}[ht]
    \caption{Methods' comparison in terms of RMSE and MAE. The reported values are averaged over all the interactions in the query set $\mathcal{E}^q$ of the test set. Bold values indicate the best method.}
    \centering
    \begin{tabular}{|c|cc|cc|cc|}
        \hline
         & \multicolumn{2}{c|}{\textbf{LiveStream-1}} & \multicolumn{2}{c|}{\textbf{LiveStream-2}} & \multicolumn{2}{c|}{\textbf{LiveStream-3}} \\ \hline
         \textbf{Baselines} & \textbf{RMSE} & \textbf{MAE} & \textbf{RMSE} & \textbf{MAE} & \textbf{RMSE}  & \textbf{MAE} \\ \hline
         \textbf{Jodie}          & $0.486 \pm 0.122$                     & $0.365 \pm 0.075$                     & $0.398 \pm 0.042$                     & $0.375 \pm 0.122$                     &   $0.492 \pm 0.016$                   & $0.426 \pm 0.023$                    \\ \hline
         \textbf{EvolveGCN}      & $0.521 \pm 0.233$                     & $0.496 \pm 0.154$                     & $0.402 \pm 0.136$                     & $0.366 \pm 0.231$                     &   $0.566 \pm 0.056$                   & $0.464 \pm 0.016$                    \\ \hline
         \textbf{DySAT}          & $0.359 \pm 0.021$                     & $0.348 \pm 0.143$                     & $0.316 \pm 0.032$                     & $0.297 \pm 0.182$                     &   $0.468 \pm 0.025$                   & $0.372 \pm 0.032$                    \\ \hline
         \textbf{VStreamDRLS}    & $0.323 \pm 0.012$                     & $0.301 \pm 0.053$                     & $0.293 \pm 0.029$                     & $0.284 \pm 0.031$                     &   $0.399 \pm 0.015$                   & $0.316 \pm 0.033$                    \\ \hline 
         \textbf{PolicyGNN}      & $0.282 \pm 0.105$                     & $0.276 \pm 0.084$                     & $0.250 \pm 0.014$                     & $0.218 \pm 0.029$                     &   $0.325 \pm 0.052$                   & $0.301 \pm 0.017$                    \\ \hline
         \textbf{MetaHIN}        & $0.274 \pm 0.021$                     & $0.269 \pm 0.013$                     & $0.242 \pm 0.058$                     & $0.204 \pm 0.027$                     &   $0.296 \pm 0.055$                   & $0.203 \pm 0.028$                    \\ \hline
         \textbf{Meta-Graph}     & $0.452 \pm 0.057$                     & $0.449 \pm 0.030$                     & $0.437 \pm 0.016$                     & $0.429 \pm 0.053$                     &   $0.373 \pm 0.026$                   & $0.297 \pm 0.018$                    \\ \hline
         \textbf{GELS}           & $0.302 \pm 0.012$         & $0.288 \pm 0.011$                                 & $0.279 \pm 0.013$                     & $0.224 \pm 0.012$                     & $0.301   \pm 0.019$                   & $0.294 \pm 0.092$                    \\ \hline
         \textbf{MELANIE-T}      & $0.286 \pm 0.032$                     & $0.238 \pm 0.012$                     & $0.251 \pm 0.106$                     & $0.246 \pm 0.37$                      &   $0.362 \pm 0.027$                   & $0.312 \pm 0.023$                    \\ \hline
         \textbf{MELANIE-B}      & $0.263 \pm 0.029$                     & $0.224 \pm 0.011$                     & $0.241 \pm 0.235$                     & $0.198 \pm 0.042$                     &   $0.355 \pm 0.010$                   & $0.295 \pm 0.028$                    \\ \hline
         \textbf{MELANIE-M}      & $0.216 \pm 0.082$                     & $0.205 \pm 0.016$                     & $0.214 \pm 0.164$                     & $0.183 \pm 0.105$                     &   $0.286 \pm 0.010$                   & $0.264 \pm 0.011$                    \\ \hline
         \textbf{MELANIE}        & \textbf{0.187} $\pm$ \textbf{0.042}   & \textbf{0.175} $\pm$ \textbf{0.026}   & \textbf{0.196} $\pm$ \textbf{0.013}   & \textbf{0.135} $\pm$ \textbf{0.052}   &   \textbf{0.223} $\pm$ \textbf{0.015} & \textbf{0.164} $\pm$ \textbf{0.011}  \\ \hline
    \end{tabular}
    \label{tab:agg_acc}
\end{table*}

\textbf{Evaluation Metrics} We evaluate the performance of our proposed model in terms of Root Mean Squared Error (RMSE) and Mean Absolute Error (MAE), which are defined as follows:

\begin{equation}
    \begin{array}{c}
         RMSE = \sqrt{\frac{1}{|\mathcal{E}^q|} \displaystyle \sum_{e^t_{u,v} \in \mathcal{E}^q} (Q(s,a) - R(s,a))^2 }  \\ \\
         MAE = \frac{\displaystyle \sum_{e^t_{u,v} \in \mathcal{E}^q} (Q(s,a) - R(s,a))^2}{|\mathcal{E}^q|}
    \end{array}
\end{equation}
where $R(s,a)$ is the reward of the throughput $b^t_{u,v}$ of the interaction $e^t_{u,v} \in \mathcal{E}^q$. Note that RMSE emphasizes more on the large errors than the MAE metric. Therefore, the RMSE metric indicates if the actions taken by the agent/tracker significantly deviate from the received reward. In addition, to measure the performance of our proposed model on how well it can adapt to new live video streaming events, we report the average reward received by the $N$ viewers for the actions taken at each $t$-th minute based on the interactions in the query set $\mathcal{E}^q$ as follows: 

\begin{equation}\label{eq:rew}
r^t = \frac{1}{N}\displaystyle \sum_{e^t_{u,v} \in \mathcal{E}^q} R(s,a)
\end{equation}

%\begin{equation}
%    r^t = \frac{1}{N}\displaystyle \sum_{e^t_{u,v} \in \mathcal{E}^q} R(s,a)
%\end{equation}

% We describe that we use three evaluation metrics. The first two of them, RMSE and MAE, measure the accuracy of the model in terms of link prediction. We mention that we exploit both MAE and RMSE, since RMSE penalizes more the large errors, compared with MAE. This means that the agent's policy cannot generalize well in this specific live video streaming event and the value of the actions taken deviate significantly against the actual reward. The third metric that we use is the cumulative reward. We mention that the scope of this metric is to evaluate how fast the agent adapts to a new environment and what is the perceived reward for the actions taken. Finally, we compare the performance of each model in terms of training time. More details will be found on supplementary.

\subsection{Compared Methods}

We evaluate the performance of the proposed MELANIE model against several graph representation learning strategies: i) \emph{Jodie}\footnote{\url{https://github.com/srijankr/jodie}}\cite{jodie2019} employs recurrent neural networks (RNNs) to update the node embeddings; ii) \emph{EvolveGCN}\footnote{\url{https://github.com/IBM/EvolveGCN}} \cite{pareja2020evolvegcn} exploits RNNs between consecutive graph convolutional networks; iii) \emph{DySAT}\footnote{\url{https://github.com/aravindsankar28/DySAT}}\cite{sankar2020dysat} adopts self-attention mechanism; and iv) \emph{VStreamDRLS}\footnote{\url{https://github.com/stefanosantaris/vstreamdrls}}\cite{antaris2020vstreamdrls} employs self-attention between consecutive graph convolutional networks. Moreover, we evaluate Melanie against the following meta-learning strategies: i) \emph{PolicyGNN}\footnote{\url{https://github.com/lhenry15/Policy-GNN}}\cite{lai2020policygnn} employs deep RL to adapt the aggregation level of convolutional networks; ii) \emph{MetaHIN}\footnote{\url{https://github.com/rootlu/MetaHIN}}\cite{Lu2020metahin} captures different semantic facets of each node in a global model; iii) \emph{Meta-Graph}\footnote{\url{https://github.com/joeybose/Meta-Graph}}\cite{bose2020metagraph} employs model-agnostic meta-learning on static networks; and iv) \emph{GELS}\footnote{\url{https://github.com/stefanosantaris/GELS}}\cite{antaris2021gels} adopts gradient boosting on live video streaming events.
To examine the different components of our model, we compare the proposed MELANIE model with the following variants: i) \emph{MELANIE-T} employs the deep RL scheme described in Section \ref{sec:proposed_model} on a single live video streaming event, without considering the meta-learning component. Moreover, MELANIE-T adopts uniform sampling in the replay memory buffer $\mathcal{D}$, instead of the prioritized sampling based on the KL-divergence. ii) \emph{MELANIE-B} incorporates the prioritized based on the KL-divergence replay memory buffer $\mathcal{D}$. Similar to the MELANIE-T model, MELANIE-B ignores the meta-learning component. iii) \emph{MELANIE-M} uses the meta-learning component in multiple events. However, MELANIE-M generates a global model, without considering the graph signatures $\mathcal{H}$ stored in the graph signature buffer $\mathcal{J}$. For reproduction purposes, the source code of the proposed MELANIE model and its variants are publicly available\footnote{\url{https://github.com/stefanosantaris/melanie}}.

% We mention that we have the graph representation learning approaches, e.g. Jodie, EvolveGCN, DySAT, VStreamDRLS. We emphasize that Jodie works on temporal interaction networks, while EvolveGCN, DySAT and VStreamDRLS are on graph snapshots.  We also mention that we compare against meta-learning approaches, such as MetaHIN, META-GRAPH and PolicyGNN. We say that we have a valinna model-agnostic meta-learning approach, namely MAML, which is a variant of MELANIE without graph signatures and without prioritization on the replay memory buffer. We also refer to our variants MELANIE-T and MELANIE-B, which are variants of MELANIE without meta-learning and with prioritized memory buffer, respectively. We explain why we selected these variants.

\textbf{Environment and Parameter Settings} We tuned the hyper-parameters of each model based on the validation set and a grid-selection strategy. We repeated our experiments five times, and the results were averaged over the five trials. In Section II of our supplementary~\cite{hyper_sup}, we detail the experimental environment and we discuss the hyper-parameter settings. Moreover, we evaluate the influence of different configurations of the replay memory and graph signature buffers on the performance of the MELANIE model.

\subsection{Performance Evaluation}

In Table \ref{tab:agg_acc}, we evaluate the performance of the examined models in terms of RMSE and MAE. The proposed MELANIE model constantly outperforms the baseline approaches in all datasets. This suggests that MELANIE can efficiently learn a policy that accurately represents the temporal interaction network. Compared with the second best method MetaHIN, the proposed MELANIE model achieves relative drops $31.8$ and $34.9\%$ in terms of RMSE and MAE in LiveStream-1, $19.0$ and $33.8\%$ in LiveStream-2, $24.6$ and $19.2\%$ in LiveStream-3. Note that the MetaHIN model outperforms the other baselines, as it employs a co-adaptation meta-learner component to capture the facets of new nodes that appear in the temporal interaction network and have a few number of interactions, as it happens in the case of live video streaming events. Instead, the other baseline approaches do not handle well the case of new nodes with few interactions in the temporal interaction network. Although MetaHIN works efficiently on a single temporal interaction network, it ignores the auxiliary information of other networks, thus having limited prediction accuracy. MELANIE overcomes this issue by performing meta-learning to generate a global model not only from a single event, but also from several past events. Moreover, we observe that MELANIE constantly outperforms the GELS baseline. Given that GELS considers equal contribution of each event during learning, it negatively impacts the model to derive the similarity between events.  In addition, on inspection of Table \ref{tab:agg_acc}, we observe that MELANIE outperforms all its variants. According to the performances of the variants we can measure the impact of the prioritized replay memory and signature buffers, as well as the meta-learning process on the link prediction accuracy of MELANIE. We find that MELANIE-B outperforms MELANIE-T, when the prioritized replay memory buffer based on the KL-divergence is employed in the learning process, demonstrating the importance of training MELANIE on divergent experiences/interactions. Moreover, MELANIE-M achieves superior performance over the variants MELANIE-B and MELANIE-T, by incorporating the information of past events via the meta-learning component. However, MELANIE beats the MELANIE-T variant, showing that the computation of the graph signatures plays a crucial role when generalizing over multiple events.

\subsection{Comparison of Meta-learning Strategies}

\begin{figure}[h] \centering

\includegraphics[scale=0.115]{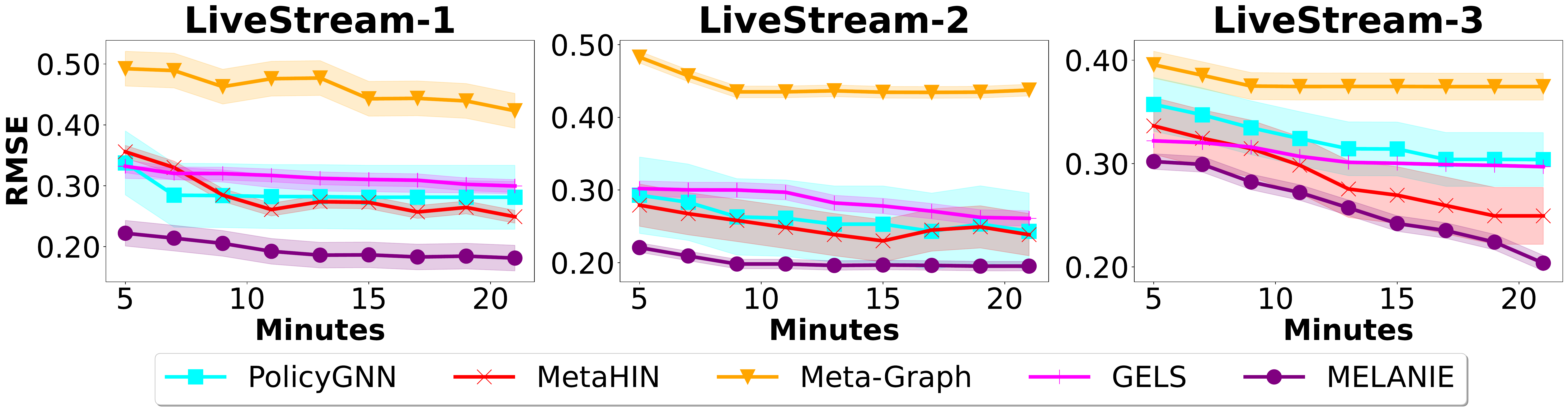}

\caption{Performance evaluation of the meta-learning approaches in terms of RMSE.} \label{fig:rmse}
\end{figure}

In Figure \ref{fig:rmse}, we present the performance of the examined meta-learning strategies in terms of RMSE during the evolution of the streaming events. We observe that MELANIE achieves a low RMSE value at the first minutes of LiveStream-1 and LiveStream-2. This occurs because each viewer in LiveStream-1 participates in more than $2$ events, as described in Section I of our supplementary~\cite{hyper_sup}. Similarly, the LiveStream-2 events occurred at a low number of offices. Therefore, the temporal interaction networks in the LiveStream-1 and LiveStream-2 datasets share structural similarities, that allows the MELANIE model to accurately learn a global policy that exhibit fast adaptation to new events in the first minutes. However, in LiveStream-3 the MELANIE model requires more streaming minutes (interactions) to adapt to the new event than in the other two datasets. This happens because the LiveStream-3  occured at $15$ different enterprise networks. This means that the temporal interaction networks in LiveStream-3 have limited structural similarities. Nevertheless, MELANIE still outperforms the baselines in the LiveStream-3 dataset, reflecting the ability of our model to better adapt to dissimilar new events than the baseline strategies. Moreover, we observe that Meta-Graph underperforms in all datasets, as Meta-Graph employs meta-learning on static graphs ignoring the evolution of the temporal interaction networks. This indicates that capturing the evolution of the network over time during the meta-learning process has a significant impact on the link prediction accuracy of the examined models.

\subsection{Average Reward Evaluation}

\begin{figure}[t] \centering
\includegraphics[scale=0.115]{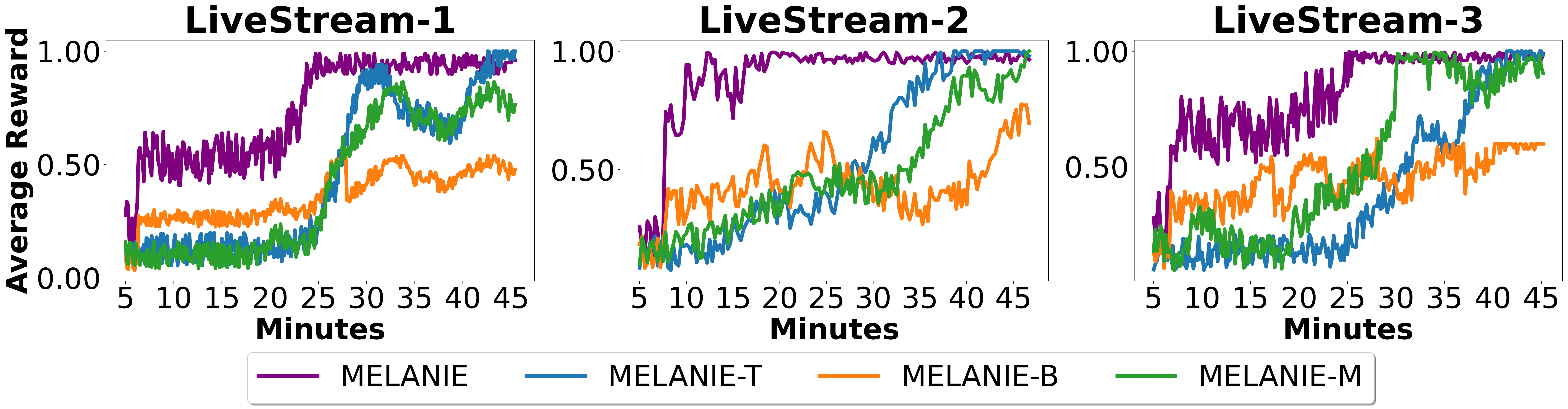}

\vspace{-0.2cm}
\caption{Average rewards (Equation~\ref{eq:rew}) of MELANIE and its variants.} \label{fig:reward}
\end{figure}

In Figure \ref{fig:reward}, we demonstrate the ability of MELANIE to achieve fast adaptation to new events based on the average reward (Equation~\ref{eq:rew}). The average reward is received by the agent/tracker as a reward for the connections that the $N$ viewers have established at each streaming minute of an event. Therefore, we consider only the approaches that employ deep RL. Note that PolicyGNN,  MELANIE and its variants are the only examined strategies that adopt deep RL. However, we omit PolicyGNN from this set of experiments, as the actions in PolicyGNN correspond to the number of convolutional layers applied on each node that participates in the network, rather than the actions to select the high-bandwidth connections between viewers. We observe that MELANIE constantly achieves higher reward than its variants in a low number of interactions (streaming minutes). This indicates the effectiveness of the meta-learning process to correctly learn a global policy that can efficiently adjust to a new event in the first streaming minutes. We also observe that the average rewards in MELANIE-T and MELANIE-M converge at a lower pace than MELANIE. This means that in MELANIE-T and MELANIE-M the agent/tracker is biased towards the low-bandwidth connections requiring a significant amount of interactions  to optimize the policy. This occurs because MELANIE-T and MELANIE-M employ uniform sampling in the replay memory buffer, instead of prioritizing the state-action transitions based on the KL-divergence, thus they are not necessarily trained on divergent interactions.

% We need to write why do we focus on the cumulative reward and why do we have such interactions. In Figure 7, we present the cummulative reward perceived by this policy for RL-based approaches only. We observe that Melanie requires less interactions with the viewers to adapt to the optimal policy that maximizes the cumulative reward. Moreover, we observe that MAML and converges in a low-value reward in few interactions in all datasets. An explanation for this can be that the agent is learning a policy that is biased towards low weights. Therefore, the connections established by this policy result in low returned rewards. In contrast, MELANIE-B exploits the prioritization of the replay memory buffer and is able to learn a better policy that results in higher rewards.

\section{Conclusions} \label{sec:conclusions}

In this study we presented the MELANIE model, a meta-RL strategy where each task corresponds a live video streaming event on large enterprise networks. We modeled each event as a MDP and then apply meta-RL to compute a global policy. To exhibit fast adaptation to a new event, we train our model on divergent interactions by prioritizing the stored state-action transitions in the replay memory buffer based on the KL-divergence of the viewers' throughputs between consecutive streaming minutes. Moreover, we introduced a graph signature buffer in the meta-learning process to measure the structural similarity among different events, allowing MELANIE to learn an accurate global model that generalizes over different events with less common viewers. Our experiments showed the proposed MELANIE model achieves high link weight prediction accuracy, with average relative drops $25.1$ and $29.3\%$ in terms of RMSE and MAE  against the second best strategy. 

Distributing a high-quality live video stream in an enterprise network is an intensive operation, as network inefficiencies in several offices negatively impact the performance of the live video streaming event. As as consequence, a low number of viewers might attend the event, resulting in limited viewer engagement \cite{Dobrian2011engagement}. Therefore, distributed live video streaming solutions require a fast and accurate selection of the high-bandwidth connections between viewers. Our model can support large enterprises to exploit the structural information from different events and efficiently distribute video content of higher quality, avoiding any network problems. An interesting future direction is to incorporate the video quality of experience perceived by each viewer in the learned policy of MELANIE, so as to maximize the user engagement \cite{Huang2018qarc}. 

\bibliographystyle{IEEEtran}
\bibliography{IEEEexample}

\end{document}